\PassOptionsToPackage{cmyk,table}{xcolor}
\documentclass[conference,pbalance]{iaria} 
\pdfoutput=1 

\usepackage[babel=true,english=american]{csquotes}
\usepackage[english]{babel} 
\usepackage[shortcuts]{extdash}

\usepackage[
  babel=true,
  expansion=alltext,
  protrusion=alltext-nott,
  nopatch=eqnum,
  final
]{microtype}

\usepackage{graphicx}
\usepackage{amsmath}
\usepackage{booktabs}
\usepackage[most]{tcolorbox}
\usepackage{orcidlink} 

\newtcolorbox{keytakeaway}[1][]{
  colback=black!4, colframe=black!55, boxrule=0.5pt, arc=2pt,
  left=5pt, right=5pt, top=4pt, bottom=4pt, boxsep=2pt,
  fonttitle=\bfseries\footnotesize, fontupper=\footnotesize,
  title={#1}, breakable
}

\usepackage{fancyhdr}

\fancypagestyle{firstpage}{
    \fancyhf{}
    \fancyhead[L]{\textit{ICSEA 2026: The Twenty-First International Conference on Software Engineering Advances}}
    
}

\title{A Closed-Loop Control Architecture for Reliable Constraint Satisfaction \\ in LLM Text Generation}

\author{%
\begin{tabular}{c@{\hspace{0.035\textwidth}}c@{\hspace{0.035\textwidth}}c}
  \begin{tabular}[t]{@{}c@{}}
    \textbf{Quan Zhou}\\
    Faculty of Information Technology\\
    and Communication Sciences\\
    Tampere University\\
    Tampere, Finland\\
    \texttt{quan.zhou@tuni.fi}
  \end{tabular}
  &
  \begin{tabular}[t]{@{}c@{}}
    \textbf{Shahbaz Siddeeq}\,*\,\orcidlink{0009-0003-9030-8841}\\
    Faculty of Information Technology\\
    and Communication Sciences\\
    Tampere University\\
    Tampere, Finland\\
    \texttt{shahbaz.siddeeq@tuni.fi}
  \end{tabular}
  &
  \begin{tabular}[t]{@{}c@{}}
    \textbf{Mika Saari}\,\orcidlink{0000-0001-7677-2355}\\
    Faculty of Information Technology\\
    and Communication Sciences\\
    Tampere University\\
    Tampere, Finland\\
    \texttt{mika.saari@tuni.fi}
  \end{tabular}
  \vspace{15pt}
\\
 
  &
  \begin{tabular}[t]{@{}c@{}}
    \textbf{Pekka Abrahamsson}\,\orcidlink{0000-0002-4360-2226}\\
    Faculty of Information Technology\\
    and Communication Sciences\\
    Tampere University\\
    Tampere, Finland\\
    \texttt{pekka.abrahamsson@tuni.fi}
  \end{tabular}
  &
   
\end{tabular}%
}

\begin{document}

\maketitle

\thispagestyle{firstpage}
\pagestyle{headings}

\begingroup
\renewcommand{\thefootnote}{*}
\footnotetext{Corresponding author: Shahbaz Siddeeq <\textit{shahbaz.siddeeq@tuni.fi}>}
\endgroup

\begin{abstract}
Software systems increasingly embed a large language model in features that must
satisfy a numeric output constraint, that is, a requirement expressible as a
number or an interval and checkable by code, such as a target word count or a
target readability grade band. Because such a model is non-deterministic, is
configured through natural-language instructions rather than a typed interface,
and satisfies a stated requirement only approximately, a single prompt neither
reliably meets the target nor preserves the source content. This paper presents
and evaluates a closed-loop control architecture for this problem. It has five
stages: generate, evaluate, adjust, archive, and analyze. The model is called
only to write and to edit text, while deterministic code compares a composite
readability value against a target band, rejects any edit that drops source
entities, numbers, or keywords, and makes every accept decision. Over 114
single-shot generation jobs and 240 closed-loop runs on four commercial models,
single-shot prompting met the target in 21.1 to 31.6 percent of cases and the
closed loop in 92.5 to 98.8 percent, within two edit rounds on average and at a
recall-based fidelity of 0.92 to 0.93; the two models common to both settings
show the same effect. Because the controller optimizes the value on which success
is scored, the result establishes reproducible control over a declared,
computable metric and not validated human difficulty. The transferable practice
is to declare the acceptance condition as code, bound the model to local edits,
and gate every edit on a content check.
\end{abstract}


\section{Introduction}\label{sec:intro}

Large Language Models (LLMs) are now ordinary
components inside software systems and are used across many software-engineering
tasks~\cite{hou2024llm4se}. Typical uses include content generation,
summarization, dialogue, and tutoring. In many of these features the surrounding
system does not merely need plausible text; it needs text that satisfies a
\emph{numeric output constraint}, that is, a requirement that the system can
express as a number or a numeric interval and check by code. Examples are an
interface string that must fit a fixed character budget, a summary that must fit
a token budget, a generated explanation that must sit at a stated reading level,
and a release note that must retain every identifier from the change log. This
paper studies the general engineering question behind these cases: how can a
system built on an LLM guarantee that a measurable property of the generated text
holds, and record why it holds?

An LLM differs from a conventional component in three ways that make such a
guarantee hard. First, it is non-deterministic: the same request can return
different text. Second, it is configured through natural-language instructions
and not through a typed interface, so there is no signature on which a numeric
requirement can be asserted and no compiler or type checker that can reject a
violation. Third, it satisfies a stated constraint only approximately, and the
error is not random but systematic, as Section~\ref{sec:results} shows. Two
further obstacles are specific to text. Numeric constraints on text interact:
making a passage simpler changes its length, and shortening a passage can
remove facts. Finally, a naive fix, namely asking the model to check its own
output, replaces one non-deterministic step with two, because the judge is the
same kind of component as the generator.

Where the constraint is not enforced, the cost falls
on people. A team either accepts silent violations, or it inserts manual review,
which removes the scaling advantage that motivated the automation. Neither option
produces an audit record, so a team cannot answer afterwards why a particular
output was shipped. Software engineering already has a standard answer for an
unreliable component: wrap it in a control loop that measures the output, checks
it against a specification, and corrects it, and keep the specification and the
decision outside the unreliable part. The engineering question is whether that
answer carries over to a component whose interface is a natural-language prompt.

This paper investigates the question through one concrete
task: generating English reading passages that meet a target word count and a
target readability band without changing the source content. The task was chosen
because it exhibits all of the properties above in a single instance. The two
constraints are numeric, they are partly in conflict, and a third constraint,
content preservation, must hold at the same time. The findings are therefore
expected to transfer to any feature with a computable target metric and a content
constraint, although this paper tests only the one application
(Section~\ref{sec:threats}).

Section~\ref{sec:related} reviews the relevant work and derives two gaps:
existing self-refinement methods keep the correctness judgment inside the model
(\textbf{G1}), and existing readability-controlled generation work characterizes
the error a prompt leaves behind rather than supplying a mechanism that removes it
(\textbf{G2}). Neither line of work reports what happens to source content while
the metric is being moved. The three research questions below each address part
of this gap.

\begin{tcolorbox}[colback=gray!2!white,colframe=black!75!black]
\textit{\textbf{RQ1.} To what extent can single-shot prompting satisfy the
length and difficulty constraints together?}
\end{tcolorbox}
\textit{Objective.} Measure how often one prompt produces text that is within the
target length and the target difficulty band. \textit{Rationale.} Single-shot
prompting is the common baseline, and prior work reports that it produces a trend
rather than reliable attainment~\cite{huang2024,hsu2025,kew2023} but does not
report joint length-and-difficulty attainment. If one prompt already satisfies
both constraints, a control loop is not needed.

\begin{tcolorbox}[colback=gray!2!white,colframe=black!75!black]
\textit{\textbf{RQ2.} How does closed-loop feedback affect difficulty control
compared with single-shot prompting?}
\end{tcolorbox}
\textit{Objective.} Measure the difficulty hit rate of the closed-loop
architecture and compare it with single-shot prompting. \textit{Rationale.} This
question targets G1 and G2. The loop adds a deterministic measurement, a bounded
edit step, and a deterministic accept check, so the accept decision no longer
depends on a model judgment.

\begin{tcolorbox}[colback=gray!2!white,colframe=black!75!black]
\textit{\textbf{RQ3.} To what extent can difficulty be controlled without
reducing the semantic fidelity of the source?}
\end{tcolorbox}
\textit{Objective.} Measure the fidelity of accepted outputs when the difficulty
hit rate is high. \textit{Rationale.} This question targets the part of the gap
that prior work leaves unmeasured. Editing text to change difficulty can remove
source content, and a controller optimizing a readability score alone has an
incentive to do so.

The purpose of this article is to specify a
reusable control architecture for LLM components under numeric output
constraints, and to report what such an architecture does and does not
demonstrate. This is a software architecture and not a new model: no model is
fine-tuned or modified, the model is used only to write and to edit text, and
every accept decision is made by deterministic code. The contributions are the
following.
\begin{itemize}
  \item A five-stage pipeline (generate, evaluate, adjust, archive, analyze)
        that turns approximate prompt constraints into measurable and
        reproducible output checks (Section~\ref{sec:arch}).
  \item A composite readability target computed from four grade-level formulas,
        with a length-compensation step that corrects a measured
        under-generation bias.
  \item A controller that applies only word and sentence replacements, guarded
        by a recall-based fidelity gate that protects source content.
  \item An evaluation against single-shot prompting over four commercial models,
        reported with the parameter settings needed to repeat it
        (Section~\ref{sec:method}), together with an analysis of control quality,
        cost, and validity, including the limit set by metric circularity
        (Sections~\ref{sec:results} and~\ref{sec:threats}).
\end{itemize}

Three limitations bound how the results should be read, and
Section~\ref{sec:threats} examines each of them. First, the controller
optimizes the same composite metric on which success is scored, so the
reported hit rate demonstrates metric controllability and not validated human
difficulty. Second, the fidelity gate is recall-based: it detects removed
content but not altered content. Third, the thresholds, the compensation
factors, and the level bands were calibrated on pilot data, and the evaluation
covers one application, one language, and four models.

The remainder of the paper is organized as follows. Section~\ref{sec:related}
reviews related work and states the two gaps. Section~\ref{sec:arch} describes
the architecture, the difficulty bands, and the fidelity gate.
Section~\ref{sec:method} describes the experimental method, the two tasks, the
generated dataset, and the settings needed to reproduce the runs.
Section~\ref{sec:results} reports the results. Section~\ref{sec:threats} states
the threats to validity. Section~\ref{sec:discussion} discusses the results and
draws the lessons for software engineers. Section~\ref{sec:conclusion} concludes
and lists future work.

\section{Related Work}\label{sec:related}

This section reviews three areas of prior work: LLMs treated as unreliable
components, readability-controlled generation, and multi-dimensional evaluation.
It closes by stating the two gaps that the research questions address.

\subsection{LLMs as Unreliable Software Components}
LLMs can be treated as unreliable components that need external checks.
Self-refinement methods, such as Self-Refine~\cite{madaan2023selfrefine} and
Reflexion~\cite{shinn2023reflexion}, let a model review and revise its own
output, and have been reported to improve output on a range of tasks. When the
model is also the judge, however, there is no external check that a target
property is met, and the judgment is itself non-deterministic: the same candidate
can be accepted in one run and rejected in the next, and no threshold is recorded
against which the decision could be replayed. The architecture in this paper
places the check outside the model, so that the same input and the same stored
state always produce the same accept or reject decision.

\subsection{Controllable and Readability-Controlled Generation}
LLMs can move text toward a target readability level, but the target is often
missed and the meaning is often changed. Huang et al.~\cite{huang2024} report
that LLMs shift educational text toward a requested level but often miss it. Hsu
et al.~\cite{hsu2025} report that readability-controlled generation produces a
trend but not stable alignment. Kew et al.~\cite{kew2023} report that
simplification behavior varies across models. Bezirhan and von
Davier~\cite{bezirhan2023} and Xiao et al.~\cite{xiao2023} generate passages and
exercises, but they use manual quality control and not a closed loop. These
results establish that a single generation step does not reliably control numeric
output properties, which is the premise of RQ1. They stop at that diagnosis: they
characterize the residual error of a prompt rather than supplying a mechanism
that removes it.

\subsection{Multi-Dimensional Evaluation}
Marulli et al.~\cite{marulli2024} report that the readability of LLM output
should not be reduced to one score. Scaria et al.~\cite{scaria2024} reach a
similar conclusion for generated questions. These results support two design
choices in this paper: a composite readability target instead of one formula, and
a fidelity gate, so that a gain on the difficulty metric cannot be obtained by
removing content.

\subsection{Gaps Addressed by This Paper}
Two gaps follow. \textbf{G1}: where a loop exists, the accept decision is made by
the model itself~\cite{madaan2023selfrefine,shinn2023reflexion}, so the loop
yields no check that the surrounding system can state, threshold, and replay.
\textbf{G2}: where a numeric text property is the object of study, the reported
result is the residual error of a prompting
strategy~\cite{huang2024,hsu2025,kew2023} rather than a mechanism that drives the
property into a declared band, and the effect of such control on source content
is not measured~\cite{marulli2024,scaria2024}. RQ1 quantifies the baseline that
G2 implies, RQ2 tests a mechanism that closes G1 and G2, and RQ3 measures the
content-preservation dimension that neither line of work reports.

\section{Architecture}\label{sec:arch}

\begin{figure*}[t]
  \centering
  \includegraphics[width=\textwidth]{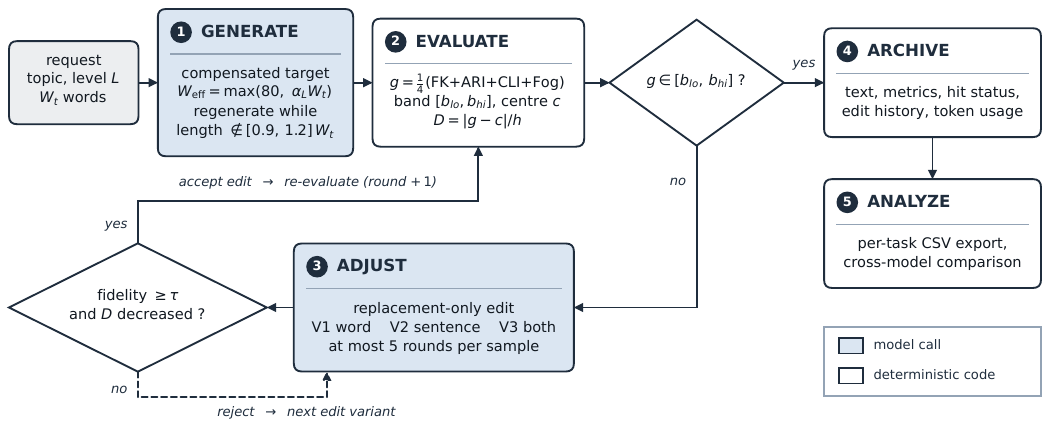}
  \caption{The five-stage closed-loop pipeline. Shaded stages call the model;
  white stages are deterministic code.}
  \label{fig:pipeline}
\end{figure*}

The architecture has five stages: generate, evaluate, adjust, archive, and
analyze. Each stage produces data that the next stage uses and that is stored for
later inspection. Figure~\ref{fig:pipeline} shows the stages and the two decision
points that close the loop. The model is called only in the generate and adjust
stages; the shaded boxes in the figure mark those two stages. Everything else is
deterministic code: metric computation, the hit check, the choice of edit
variant, the fidelity check, and the accept decision. A candidate leaves the loop
and is archived only when it is inside the difficulty band. Otherwise the
controller selects an edit variant, and the edited candidate is accepted back
into the loop only if it passes the fidelity gate and has moved closer to the
band centre; if it does not, the controller discards it and tries another
variant. The loop terminates on a hit or after five adjustment rounds. This
design confines non-determinism to two stages and makes every accept or reject
decision reproducible from the stored state.

\subsection{Generate Stage and Length Compensation}
Pilot runs indicated that prompts tended to produce text that was too short, and
that the shortfall grew as the requested difficulty level fell. The generate
stage corrects this bias before the difficulty loop starts. For a user target of
$W_t$ words at level $L$, the prompt requests a compensated target
\begin{equation}
W_{\mathrm{eff}} \;=\; \max\!\left(80,\ \operatorname{round}(\alpha_L \cdot W_t)\right),
\end{equation}
where $\alpha_L$ is the level-specific compensation factor listed in
Table~\ref{tab:levels}. Candidates are still scored against the user's original
target $W_t$: a candidate is accepted if its length lies in
$[0.9\,W_t,\ 1.2\,W_t]$. If it does not, the system generates again and keeps the
candidate whose length is closest to $W_t$. The compensation is therefore a
generation-stage calibration and not a relaxation of the evaluation standard.

\subsection{Difficulty Levels, Target Bands, and the Hit Condition}\label{sec:target}
Difficulty is expressed as one of five discrete levels, $L\in\{1,\dots,5\}$, where
Level~1 is the least complex and Level~5 the most complex. A level is not a
label given to the model to interpret; it is a numeric interval over a computed
value. The value is the \emph{grade-family mean} $g$, the mean of four
grade-level readability indices: Flesch-Kincaid Grade (FK)~\cite{kincaid1975},
Automated Readability Index (ARI)~\cite{smith1967}, Coleman-Liau Index
(CLI)~\cite{coleman1975}, and Gunning Fog~\cite{gunning1952}. Flesch Reading
Ease~\cite{flesch1948} is computed for monitoring but takes no part in the
decision. Each level $L$ maps to a band
$[\,b_{\text{lo}},\,b_{\text{hi}}\,]$ on the US grade scale, given in
Table~\ref{tab:levels}, and the hit condition is
\begin{equation}
\text{hit} \;=\; \big(b_{\text{lo}} \le g \le b_{\text{hi}}\big),
\quad
g = \tfrac{1}{4}\!\left(\text{FK}+\text{ARI}+\text{CLI}+\text{Fog}\right).
\end{equation}
The bands were calibrated against common grade-band representations for English
reading material and against the readability distribution observed in pilot runs.
The four formulas are correlated but not identical, so requiring each one to fall
inside a narrow band is frequently unsatisfiable; the mean is a single value that
the controller can move. A band is used instead of a point because reading
material covers a range of complexity and not one exact value. Two derived
quantities are used throughout the paper. For a band with centre
$c=(b_{\text{lo}}+b_{\text{hi}})/2$ and half-width
$h=(b_{\text{hi}}-b_{\text{lo}})/2$, the normalized distance is $D=|g-c|/h$, and
the \emph{grade drift} of a sample is the signed quantity $g-c$, reported in
grade points. Drift therefore states both the size and the direction of the
error: a positive value means the text is harder than the band centre.

\begin{table}[t]
\caption{Difficulty levels: target band on the grade-family mean $g$ and the
length-compensation factor $\alpha_L$ used by the generate stage.}
\label{tab:levels}
\centering
\begin{tabular}{lccc}
\toprule
Level & Band $[b_{\text{lo}},b_{\text{hi}}]$ & Centre $c$ & $\alpha_L$ \\
\midrule
1 & 6--8   & 7.0  & 1.40 \\
2 & 8--10  & 9.0  & 1.30 \\
3 & 10--12 & 11.0 & 1.05 \\
4 & 12--14 & 13.0 & 1.00 \\
5 & 14--20 & 17.0 & 1.00 \\
\bottomrule
\end{tabular}
\end{table}

\subsection{Closed-Loop Edit Controller}
The controller does not regenerate the full text in each round. It applies only
replacements. There are three variants: word replacement (V1),
sentence-structure change (V2), and both together (V3). The controller selects V1
when the mismatch is lexical, V2 when the mismatch is structural, and V3 when
both need to change. After each edit the system recomputes the metrics, rechecks
the hit condition, and recomputes $D$. An edited candidate replaces the current
text only if it passes the fidelity gate and either hits the band or reduces $D$;
otherwise it is discarded and another variant is tried. A bounded edit reduces
cost and reduces the risk of changing meaning, because each round changes only
what the diagnosis requires.

\subsection{Fidelity Gate}\label{sec:fidelity}
Each candidate edit passes a fidelity gate before it is accepted. The gate
extracts the source entities, numbers, and top keywords and computes their recall
in the candidate. The fidelity score is a weighted sum of three recall values:
entity recall ($R_e$), number recall ($R_n$), and keyword recall ($R_k$):
\begin{equation}
\textit{Fidelity} \;=\; 0.45\,R_e + 0.35\,R_n + 0.20\,R_k .
\end{equation}
Entities and numbers carry the facts, so they have higher weight. A candidate is
eligible for acceptance if its fidelity score is at least $\tau=0.72$. Some local
lexical or syntactic acceptances require at least 0.85. These thresholds were set
on pilot data to balance the acceptance rate against content loss. They are
parameters and not validated constants (Section~\ref{sec:threats}).

\subsection{Archive and Analyze}
The archive layer stores each generated or edited text with its readability
metrics, experiment identifiers, request and generation metadata, hit status, the
edit history, and token usage. The analyze layer exports one comma-separated file
per task. These files support repeated cross-model comparison and let any
controller decision be traced back to the state that produced it. This record is
what makes the system auditable: for any shipped output, the stored state names
the metric value, the band, the fidelity score, and the edit that was applied.

\section{Experimental Method}\label{sec:method}
This section restates the research questions in operational terms, defines the
two tasks and the dataset they produced, lists the metrics, and gives the
settings needed to repeat the runs.

\subsection{Research Questions and Hypotheses}
Each research question is answered by one task. RQ1 is answered by Task~A,
using hit rate and length deviation; RQ2 by Task~B, using the difficulty hit
rate, together with the two models that appear in both tasks; RQ3 by Task~B,
using the fidelity score. The success criteria are stated as hypotheses. H1:
single-shot prompting produces fluent text but does not satisfy the difficulty
target reliably. H2: closed-loop feedback raises the difficulty hit rate above
90 percent. H3: the average recall-based fidelity stays at or above 0.85.

\subsection{The Two Tasks}
The two tasks correspond to two stages of the same workflow, and they isolate the
contribution of the closed loop.

Task~A, the baseline, measures what one prompt achieves with no feedback. The
system is given a topic, a difficulty level, and a target word count, and it
produces a passage in a single generation call. The length-compensation step
of Section~\ref{sec:arch} is active, so Task~A is a fair rather than a
weakened baseline; the difficulty loop is not. A sample is a hit if the
grade-family mean of that one passage falls inside the band of the requested
level. Task~A therefore answers whether the closed loop is needed at all.

Task~B measures the full architecture. The system is given an existing passage,
drawn from the archive of previously generated texts, together with a target
level, and it runs the
evaluate--adjust--gate loop of Figure~\ref{fig:pipeline} until the passage is
inside the band or the round limit is reached. A sample is a hit under the same
condition as in Task~A. The two tasks therefore differ in their starting point as
well as in the loop: Task~A controls from scratch, Task~B from a given text
(Section~\ref{sec:threats}).

\subsection{Dataset}
No external corpus is used. The evaluation set is generated by the system itself
and is fully specified by the configuration below, which is what makes it
reproducible.

For Task~A, the configuration enumerates four topic prompts
(\emph{climate adaptation}, \emph{public transportation}, \emph{ancient trade
routes}, and \emph{marine ecosystems}), five difficulty levels, and two target
lengths (300 and 600 words). This gives $4\times5\times2=40$ planned generation
jobs per model. The reported set contains 38 jobs per model, or 114 in total. Two
jobs per model were excluded because generation or transfer failed and no text
was returned; these were not low-scoring outputs, so the exclusion does not
favour the baseline or the loop. For Task~B, source passages are drawn from the
archive produced by the generation runs and are adapted to target levels 1 to 5.
The reported set contains 240 adaptation runs, 80 per model.

\subsection{Metrics}
Task~A is measured by hit rate, length deviation, and average attempts. Task~B is
measured by hit rate, residual drift, average rounds, fidelity score, and token
consumption. Hit rate is the percentage of samples satisfying the hit condition
of Section~\ref{sec:target}. Length deviation is the mean absolute percentage
difference between the produced word count and the user target $W_t$. Average
attempts counts generation calls per Task~A sample, including regenerations
triggered by the length window. Average rounds counts edit rounds per Task~B
sample. Residual drift is the signed grade drift $g-c$ remaining after
adjustment, in grade points. Fidelity is defined in Section~\ref{sec:fidelity}.
Token consumption is the total prompt and completion tokens charged per final
accepted sample, summed over every call the loop made for that sample.

\subsection{Models, Settings, and Reproducibility}
The two tasks ran as separate campaigns. Task~A used gpt-5-mini,
grok-4.1-fast, and gemini-3-flash-preview (written
gemini-3-flash below). Task~B used
gpt-5-mini, grok-4.1-fast, and deepseek-v3.2. All
models were accessed as hosted commercial endpoints through a single
OpenRouter-compatible gateway, so that only the model identifier changed between
campaigns. Two models, gpt-5-mini and grok-4.1-fast, appear in
both campaigns and therefore support a within-model comparison of the two
settings.

The controller parameters are those stated in Section~\ref{sec:arch} and are
repeated here so that a run can be reconstructed: the level bands and
compensation factors of Table~\ref{tab:levels}; the length acceptance window
$[0.9\,W_t,\ 1.2\,W_t]$; the composite $g$ over FK, ARI, CLI, and Fog with equal
weights; the fidelity weights $0.45/0.35/0.20$; the fidelity threshold
$\tau=0.72$, raised to 0.85 for local lexical and syntactic acceptances; the
local-improvement criterion $D=|g-c|/h$ must decrease; and a hard limit of five
adjustment rounds per sample. Experiments were executed by configuration-driven
batch scripts that call the same service endpoints as the interactive interface,
rather than through manual interaction, so that every sample carries an
experiment identifier, a batch identifier, and a sample identifier derived from
its topic, level, target length, and repeat index. Readability metrics are
computed by deterministic code from the stored text, so any reported metric can
be recomputed from the archive without calling a model again.

Because the model sets differ between campaigns and the Task~A sample is small,
all cross-model differences are reported as descriptive and no significance test
is run.

\section{Results}\label{sec:results}
This section reports Task~A, then Task~B, then the within-model comparison, and
finally token cost and residual drift.

\subsection{RQ1: Single-Shot Prompting Does Not Meet the Difficulty Target}
Table~\ref{tab:taskA} reports Task~A. Length control was the easier of the two
constraints: mean length deviation was 8.11 percent for gpt-5-mini,
7.46 percent for grok-4.1-fast, and 11.89 percent for
gemini-3-flash, that is, within roughly one tenth of the
requested word count. Difficulty control was not. The hit rate was 31.6 percent
for gpt-5-mini, 21.1 percent for grok-4.1-fast, and 31.6
percent for gemini-3-flash, so between two thirds and four
fifths of the passages fell outside the requested band. Average attempts ranged
from 1.84 to 2.05, which means the length-retry mechanism was exercised on
roughly every second sample and still left the difficulty error in place. A
correct length did not imply a correct difficulty.

\begin{table}[t]
\caption{Task~A: single-shot generation, 38 samples per model.}
\label{tab:taskA}
\centering
\resizebox{\columnwidth}{!}{%
\begin{tabular}{lccc}
\toprule
Model & Length dev.\ (\%) & Hit rate (\%) & Avg.\ attempts \\
\midrule
gpt-5-mini             & 8.11  & 31.6 & 1.84 \\
grok-4.1-fast          & 7.46  & 21.1 & 2.05 \\
gemini-3-flash & 11.89 & 31.6 & 1.97 \\
\bottomrule
\end{tabular}
}
\end{table}

\begin{figure}[t]
  \centering
  \includegraphics[width=\columnwidth]{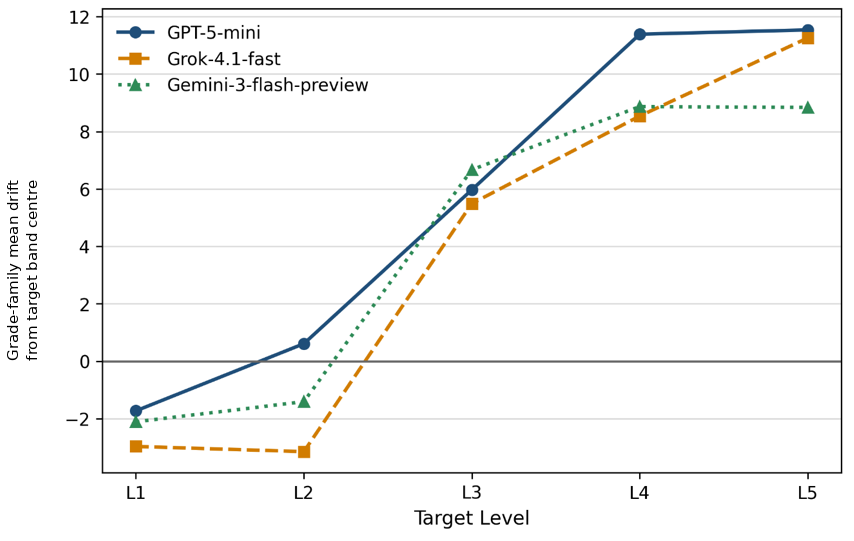}
  \caption{Single-shot prompting (Task~A): grade drift $g-c$ from the target band
  centre, by requested level. Positive means harder than requested.}
  \label{fig:drift}
\end{figure}

The error is systematic rather than random, which is what makes it correctable.
Figure~\ref{fig:drift} breaks the error down by requested level and plots the
grade drift $g-c$ defined in Section~\ref{sec:target}. All three models undershoot
at the two easiest levels, by about $-1.7$ to $-3.2$ grade points at Levels~1
and~2, and overshoot from Level~3 upward, by about $+5.5$ to $+6.7$ grade points
at Level~3 and by $+8.6$ to $+11.6$ grade points at Levels~4 and~5. The three
curves have the same shape, so the bias is a property of prompting for a level
rather than of one vendor. A model that misses the requested level this
consistently cannot produce level-specific material by prompt alone, and the
regularity of the curve is precisely the signal a controller can act on: the sign
of the drift names the direction of the required edit.

\begin{keytakeaway}[Key takeaway, RQ1]
Single-shot prompting does not control numeric output properties reliably. It
produces fluent text of roughly the right length but misses the difficulty band
in 68 to 79 percent of samples. The difficulty error is systematic: too easy at
Levels~1 and~2, too hard at Levels~3 to~5. What is needed is a feedback
mechanism, not a better prompt.
\end{keytakeaway}

\subsection{RQ2: Closed-Loop Control Meets the Target}
\begin{figure}[t]
  \centering
  \includegraphics[width=\columnwidth]{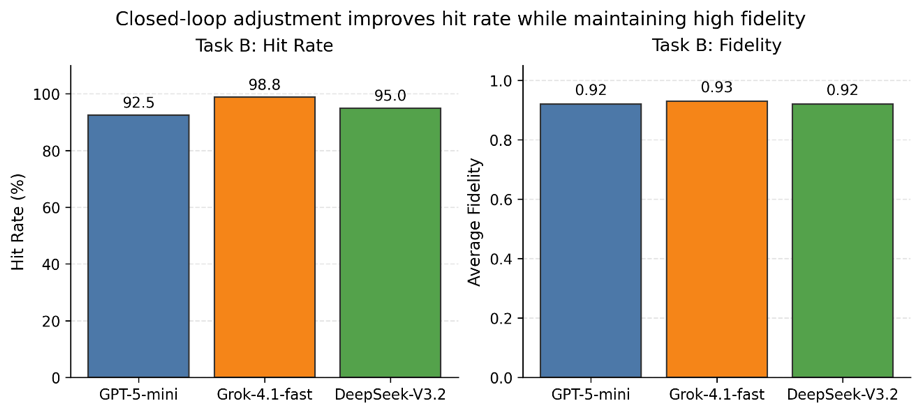}
  \caption{Task~B: closed-loop hit rate (left) and average fidelity (right) by
  model. Hit rate rises while fidelity stays at 0.92 to 0.93.}
  \label{fig:hitrate}
\end{figure}

Table~\ref{tab:taskB} and Figure~\ref{fig:hitrate} report Task~B. The hit rate
was 92.5 percent for gpt-5-mini, 98.8 percent for
grok-4.1-fast, and 95.0 percent for deepseek-v3.2, so all three
models are above the 90 percent threshold of H2. The average number of edit
rounds was 1.68 to 1.78, that is, below two and far below the limit of five. The
two numbers should be read together: a high hit rate reached after many rounds
would indicate that the loop was simply resampling until something passed,
whereas a high hit rate reached in fewer than two rounds indicates that the
diagnosis, the choice of edit variant, and the accept check are doing the work.
Where prior readability-control studies report a trend but not stable
attainment~\cite{huang2024,hsu2025}, the loop reaches the declared band.

\begin{table}[t]
\caption{Task~B: closed-loop architecture, 240 samples (80 per model).}
\label{tab:taskB}
\centering
\resizebox{\columnwidth}{!}{%
\begin{tabular}{lccc}
\toprule
Model & Hit rate (\%) & Avg.\ rounds & Avg.\ fidelity \\
\midrule
gpt-5-mini    & 92.5 & 1.78 & 0.92 \\
grok-4.1-fast & 98.8 & 1.74 & 0.93 \\
deepseek-v3.2 & 95.0 & 1.68 & 0.92 \\
\bottomrule
\end{tabular}
}
\end{table}

\subsection{The Effect Is Also Observed Within Model}
The two campaigns do not use identical model sets, so part of the aggregate jump
could in principle reflect the change of model rather than the change of
architecture. Two models, however, appear in both campaigns, and for those the
comparison is within model: gpt-5-mini moves from 31.6 to 92.5 percent,
and grok-4.1-fast moves from 21.1 to 98.8 percent. The effect is
therefore not carried by gemini-3-flash, which appears only in
the baseline, or by deepseek-v3.2, which appears only in the closed
loop; the two models common to both settings show the largest and the
second-largest gain in the study. This is a within-model comparison of two
settings and not a controlled ablation, because Task~A generates from scratch and
Task~B adapts an existing passage (Section~\ref{sec:threats}).

\begin{keytakeaway}[Key takeaway, RQ2]
The control loop moves the optimized metric into the target band. The hit rate
rose from at most 31.6 percent under single-shot prompting to 92.5 to 98.8
percent, and the same direction and magnitude of change is observed within the
two models present in both settings. The average number of rounds was below two,
so the result comes from the deterministic evaluate, adjust, and accept logic and
not from many retries. The result establishes metric controllability
(Section~\ref{sec:threats}).
\end{keytakeaway}

\subsection{RQ3: Recall-Based Fidelity Is Kept}
Average fidelity was 0.92 to 0.93 across the three models while the hit rate rose
(Figure~\ref{fig:hitrate}), against a gate threshold of $\tau=0.72$ and an H3
threshold of 0.85. The margin matters for the interpretation: accepted outputs
did not merely clear the gate, they cleared it by roughly 0.20, so the controller
was not trading content away to the last admissible point in order to reach the
band. Read against the failure mode this metric was designed to catch, the result
says that the gain in difficulty control was not obtained by deleting source
entities, numbers, or keywords. It says nothing about content that was kept but
altered, because the score is recall-based; Section~\ref{sec:threats} sets out
what that leaves open.

\begin{keytakeaway}[Key takeaway, RQ3]
The fidelity gate lets the loop change readability and keep the source content.
Average fidelity stayed at 0.92 to 0.93, about 0.20 above the acceptance
threshold, while the hit rate rose. Difficulty control and content preservation
held together. The check is recall-based: it confirms that source entities,
numbers, and keywords are kept, not that every fact is unchanged
(Section~\ref{sec:threats}).
\end{keytakeaway}

\subsection{Cost and Residual Drift}
The token cost per final accepted sample was 8{,}754 for
deepseek-v3.2, 15{,}949 for grok-4.1-fast, and 19{,}605 for
gpt-5-mini, a spread of about 2.2 times between the cheapest and the most
expensive. The choice of model is therefore a deployment trade-off rather than an
architectural one: grok-4.1-fast had the highest hit rate, while
deepseek-v3.2 reached a comparable hit rate and the same fidelity at
about 55 percent of the token cost of grok-4.1-fast and about 45
percent of that of gpt-5-mini.

\begin{figure}[t]
  \centering
  \includegraphics[width=\columnwidth]{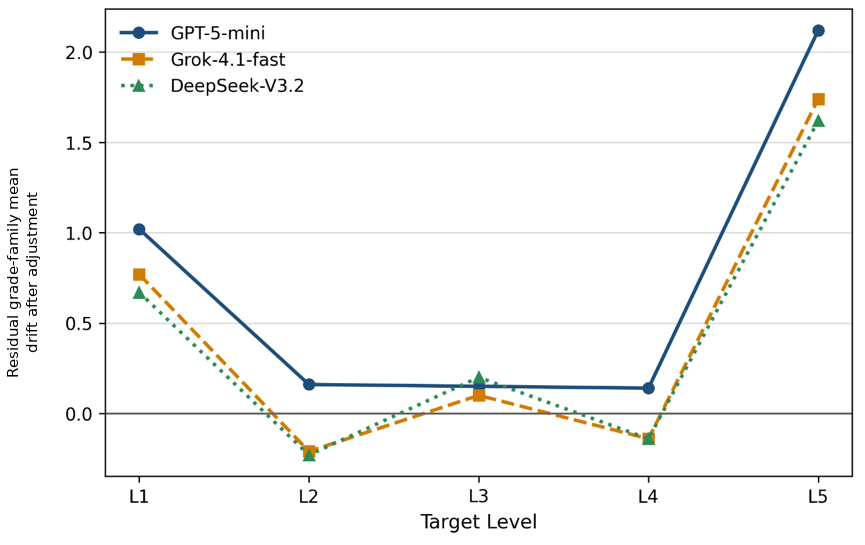}
  \caption{Task~B: residual grade drift after adjustment, by target level. The
  drift is near zero for Levels~2 to~4 and positive at Level~5 for all models.}
  \label{fig:residual}
\end{figure}

Figure~\ref{fig:residual} reports the signed drift that remains after
adjustment. For Levels~2 to~4 it is close to zero for all three models,
between $-0.23$ and $+0.19$ grade points, that is, well inside the band. At
Level~1 a positive drift of 0.66 to 1.02 remains, which matters because that
band is the narrowest in the study (half-width 1.0), so the residual sits at
or beyond the band edge. At Level~5 the drift stays positive for every model,
at 1.61 to 2.12 grade points, inside the wider half-width of 3.0 but
one-sided. Two causes are plausible and cannot be separated with the present
data: Level~5 sits at the top of the scale, where the readability formulas
respond sharply to denser vocabulary and longer clauses, and the
replacement-only controller is deliberately conservative, so it has little
room to move a text that must be highly advanced yet still centred.

Control is therefore tightest in the middle of the scale, where the controller
has room on both sides of the band, and it degrades at the extremes, where it
does not. The Level~5 residual is small in absolute terms but one-sided, and a
one-sided error is the kind that a longer run will not average away. It bounds
the claim: the architecture controls the composite metric reliably at Levels~2
to~4 and with a known positive bias at Level~5.

\section{Threats to Validity}\label{sec:threats}
\textbf{Construct validity: metric circularity.} The controller optimizes the
composite mean $g$, and Task~B scores success by whether the same $g$ falls in
the target band, so the loop edits until the success criterion is met. The 92.5
to 98.8 percent figures therefore measure whether a declared, computable quantity
can be driven into a declared interval and the decision kept auditable, which is
an engineering property; as a statement about text difficulty they are close to
tautological. Readability formulas use surface features, such as sentence length
and word complexity, and do not measure discourse structure, concept density, or
reader background. The question this design cannot answer is whether the output
reads as genuinely easier or harder to a person, and answering it needs an
independent, non-optimized criterion, that is, human or expert judgment on the
accepted outputs. Two things limit the damage. The circularity is a property of
the evaluation and not of the mechanism: the same loop accepts any computable
acceptance predicate, including one supplied by an external judge. Moreover, the
criterion is not satisfied by construction, since 1.2 to 7.5 percent of Task~B
samples never reach the band within the round limit and the residual drift of
Figure~\ref{fig:residual} stays non-zero at Levels~1 and~5.

\textbf{Construct validity: recall-only fidelity.} The fidelity score is
recall-based: it detects the removal of source entities, numbers, and keywords,
but not a changed or invented fact. A passage that rewrites ``increased by 20
percent'' as ``decreased by 20 percent'' preserves both the entity and the number
and would pass. The reported 0.92 to 0.93 therefore supports the narrower claim
that accepted edits retained the source entities, numbers, and keywords, which is
the failure mode a readability-optimizing controller is most tempted by, since
deleting a difficult named entity or a long number is the cheapest way to move a
grade score. It does not support a claim that the facts are unchanged, and the
two claims should not be conflated. Adding a factual-consistency check, for
example, one based on natural-language inference, would change the predicate
evaluated at the gate and not the architecture.

\textbf{Internal validity.} The fidelity thresholds (0.72 and 0.85), the
length-compensation factors, the level bands, and the round limit were set on
pilot data. They are parameters and not validated constants, and they may not
transfer to other domains, languages, or models. Their values are listed in
Section~\ref{sec:method} so that a replication can vary them.

\textbf{Conclusion validity.} The Task~A reported set is small, at 38 samples per
model, and no significance test is run, so all cross-model differences are
descriptive. The two settings also differ in more than the presence of the loop:
Task~A generates a passage from scratch, whereas Task~B adapts an existing
passage, so the comparison isolates the value of a closed loop over a one-shot
pipeline rather than the value of the loop with every other factor held fixed.
Two models appear in both settings and show the same direction and magnitude of
change, which reduces but does not remove this concern. The comparison rests on
the size of the difference, roughly threefold to fivefold, rather than on a
statistical test.

\textbf{External validity.} The evaluation uses one application (English reading
passages), one language, and four models; other content types, languages, and
models are not tested. The architecture is not specific to this application: in
principle, any task with a computable target metric and a content constraint
could use the same loop, although this was not tested here.

\section{Discussion}\label{sec:discussion}
The results support a narrow claim and suggest a broader one. The narrow claim is
that a numeric property of LLM output can be driven into a declared interval
reliably, cheaply, and auditably: 92.5 to 98.8 percent attainment at fewer than
two edit rounds, against 21.1 to 31.6 percent for the prompt alone. The broader
suggestion is that the reason this works has little to do with reading passages.
It works because the loop moves the definition of correctness out of the prompt
and into code. Once correctness is a predicate over a computed value, the
non-deterministic component is no longer being asked to decide anything; it is
being asked to produce a candidate that code will accept or reject.

One result qualifies this on the design side: control degrades at the ends of the
scale (Figure~\ref{fig:residual}), which suggests that a bounded,
replacement-only controller needs slack on both sides of its target, and that a
system with hard limits at the extremes should expect a one-sided residual error
there.

The architecture generalizes to any feature in which an LLM must satisfy a
requirement that code can evaluate, and five lessons follow for practitioners.
\begin{itemize}
  \item L1: make the acceptance criterion code and not prose. A prompt
        requirement (``about 300 words, at a fifth-grade level'') is not
        testable, whereas the same requirement expressed as a predicate over a
        computed value is testable, loggable, and replayable. This is the step
        that converts an LLM feature from something a team hopes about into
        something a team can put a test around.
  \item L2: keep the judge out of the model. Self-evaluation reproduces the
        component's non-determinism inside the accept decision, so an output
        cannot be explained after the fact. Here the model appears in two of five
        stages and the three deterministic stages own every decision, which is
        what closes G1 and what makes a stored decision replayable.
  \item L3: prefer a composite target over a single score, and accept a band
        rather than a point. Four correlated readability formulas rarely agree
        inside a narrow interval, so requiring all four made the objective
        frequently unsatisfiable, whereas the mean gave the controller a single
        quantity to move (Section~\ref{sec:target}). The same holds for any
        feature with several correlated quality signals.
  \item L4: bound the edit, and gate it on what must not change.
        Replacement-only edits kept the loop at fewer than two rounds and
        fidelity at 0.92 to 0.93, whereas full regeneration would have discarded
        the previous round's work each time. Pair every optimization target with
        a preservation constraint, because a controller scored on one metric will
        otherwise reach it by damaging something the metric does not see.
  \item L5: measure the cost of control, and treat the model as a deployment
        parameter. The loop costs one to two extra model calls per sample, and
        the token cost per accepted sample varied by a factor of 2.2 across
        models at comparable quality; because the architecture holds the model at
        arm's length, that ratio is a deployment choice and not a redesign.
\end{itemize}

A final lesson concerns reporting rather than design. A closed loop built on a
self-optimized metric reports how well it satisfies itself. That is a real
engineering result, because a check that holds and is logged is worth more than an
instruction that usually works, but it is not evidence about end users. Such a
system should say so in the abstract and not only in the threats section, and
should separate the evidence it has, namely that the metric is controllable and
the content check holds, from the interpretation a reader will otherwise supply,
namely that the text is genuinely easier to read. The practical rule is to add an
independent check, drawn from outside the optimization loop, before claiming
anything about end users; in this architecture that check is cheap to add,
because the gate is a pluggable predicate, but it is not already done.

\section{Conclusion and Future Work}\label{sec:conclusion}
This paper presented a closed-loop architecture that turns approximate prompt
constraints into measurable and reproducible output checks. Single-shot prompting
met the difficulty target in at most 31.6 percent of cases, with a systematic
rather than a random error (RQ1). The closed-loop architecture met it in 92.5 to
98.8 percent of cases at fewer than two edit rounds on average, a change that
also holds within the two models present in both settings (RQ2), while
recall-based fidelity stayed at 0.92 to 0.93 (RQ3). The evidence establishes
metric controllability and not validated human difficulty, and it is bounded by
the recall-only fidelity score, the one-sided residual drift at Level~5, the
pilot-calibrated parameters, and the descriptive statistics. For practitioners
the transferable result is the five lessons of Section~\ref{sec:discussion}, of
which the first is load-bearing: an LLM feature becomes engineerable at the
moment its acceptance criterion is written as code rather than as prose.

Future work follows from those limits: human and expert evaluation, to test
perceived difficulty against an independent criterion; a factual-consistency
check, to detect altered and not only removed content; a fully model-controlled
comparison, with the same model set and the same starting texts in both settings
and enough samples for a significance test, to separate the effect of the loop
from that of the task shape; adaptive step size or an explicit overshoot penalty,
to improve control at the top band; and application to other domains, languages,
and constraint types, to test whether the pattern and not only the result
transfers.

\section*{Acknowledgment}
This work has been supported by FAST, the Finnish Software Engineering Doctoral Research Network, funded by the Ministry of Education and Culture, Finland.

\section*{Declaration of AI Assistance}
During the preparation of this manuscript, the authors used ChatGPT to assist with grammar refinement, sentence restructuring, and formatting improvements. Following the
use of this tool, the authors carefully reviewed and revised the content and assume full responsibility for the final version of the publication.

\printbibliography

@inproceedings{huang2024,
  author    = {Huang, Chieh-Yang and Wei, Jing and Huang, Ting-Hao Kenneth},
  title     = {Generating Educational Materials with Different Levels of Readability Using {LLMs}},
  booktitle = {Proc. 3rd Workshop on Intelligent and Interactive Writing Assistants (In2Writing '24)},
  pages     = {16--22},
  year      = {2024},
  doi       = {10.1145/3690712.3690718}
}

@inproceedings{hsu2025,
  author    = {Hsu, Yi-Sheng and Feldhus, Nils and Hakimov, Sherzod},
  title     = {Free-Text Rationale Generation under Readability Level Control},
  booktitle = {Proc. 4th Workshop on Generation, Evaluation and Metrics (GEM 2025)},
  pages     = {129--150},
  year      = {2025},
  url       = {https://aclanthology.org/2025.gem-1.11/}
}

@inproceedings{kew2023,
  author    = {Kew, Tannon and Chi, Alison and V{\'a}squez-Rodr{\'i}guez, Laura and Agrawal, Sweta and Aumiller, Dennis and Alva-Manchego, Fernando and Shardlow, Matthew},
  title     = {{BLESS}: Benchmarking Large Language Models on Sentence Simplification},
  booktitle = {Proc. 2023 Conference on Empirical Methods in Natural Language Processing (EMNLP)},
  pages     = {13291--13309},
  year      = {2023},
  doi       = {10.18653/v1/2023.emnlp-main.821}
}

@article{bezirhan2023,
  author    = {Bezirhan, Ummugul and von Davier, Matthias},
  title     = {Automated Reading Passage Generation with {OpenAI}'s Large Language Model},
  journal   = {Computers and Education: Artificial Intelligence},
  volume    = {5},
  pages     = {100161},
  year      = {2023},
  doi       = {10.1016/j.caeai.2023.100161}
}

@inproceedings{xiao2023,
  author    = {Xiao, Changrong and Xu, Sean Xin and Zhang, Kunpeng and Wang, Yufang and Xia, Lei},
  title     = {Evaluating Reading Comprehension Exercises Generated by {LLMs}: A Showcase of {ChatGPT} in Education Applications},
  booktitle = {Proc. 18th Workshop on Innovative Use of NLP for Building Educational Applications (BEA 2023)},
  pages     = {610--625},
  year      = {2023},
  doi       = {10.18653/v1/2023.bea-1.52}
}

@article{marulli2024,
  author    = {Marulli, Fiammetta and Campanile, Lelio and de Biase, Maria Stella and Marrone, Stefano and Verde, Laura and Bifulco, Marco},
  title     = {Understanding Readability of Large Language Models Output: An Empirical Analysis},
  journal   = {Procedia Computer Science},
  volume    = {246},
  pages     = {5273--5282},
  year      = {2024},
  doi       = {10.1016/j.procs.2024.09.636}
}

@incollection{scaria2024,
  author    = {Scaria, Nicy and Chenna, Suman Dharani and Subramani, Deepak},
  title     = {Automated Educational Question Generation at Different {Bloom}'s Skill Levels Using Large Language Models: Strategies and Evaluation},
  booktitle = {Artificial Intelligence in Education},
  series    = {Lecture Notes in Computer Science},
  volume    = {14830},
  pages     = {165--179},
  publisher = {Springer},
  year      = {2024},
  doi       = {10.1007/978-3-031-64299-9_12}
}

@inproceedings{madaan2023selfrefine,
  author    = {Madaan, Aman and Tandon, Niket and Gupta, Prakhar and Hallinan, Skyler and Gao, Luyu and Wiegreffe, Sarah and Alon, Uri and Dziri, Nouha and Prabhumoye, Shrimai and Yang, Yiming and Gupta, Shashank and Majumder, Bodhisattwa Prasad and Hermann, Katherine and Welleck, Sean and Yazdanbakhsh, Amir and Clark, Peter},
  title     = {Self-Refine: Iterative Refinement with Self-Feedback},
  booktitle = {Advances in Neural Information Processing Systems (NeurIPS)},
  volume    = {36},
  year      = {2023}
}

@article{shinn2023reflexion,
  title={Reflexion: Language agents with verbal reinforcement learning},
  author={Shinn, Noah and Cassano, Federico and Gopinath, Ashwin and Narasimhan, Karthik and Yao, Shunyu},
  journal={Advances in neural information processing systems},
  volume={36},
  pages={8634--8652},
  year={2023}
}

@article{hou2024llm4se,
  title={Large language models for software engineering: A systematic literature review},
  author={Hou, Xinyi and Zhao, Yanjie and Liu, Yue and Yang, Zhou and Wang, Kailong and Li, Li and Luo, Xiapu and Lo, David and Grundy, John and Wang, Haoyu},
  journal={ACM Transactions on Software Engineering and Methodology},
  volume={33},
  number={8},
  pages={1--79},
  year={2024},
  publisher={ACM New York, NY}
}

@article{flesch1948,
  author  = {Flesch, Rudolph},
  title   = {A New Readability Yardstick},
  journal = {Journal of Applied Psychology},
  volume  = {32},
  number  = {3},
  pages   = {221--233},
  year    = {1948}
}

@techreport{kincaid1975,
  author      = {Kincaid, J. Peter and Fishburne, Robert P. and Rogers, Richard L. and Chissom, Brad S.},
  title       = {Derivation of New Readability Formulas for Navy Enlisted Personnel},
  institution = {Naval Technical Training Command},
  number      = {Research Branch Report 8-75},
  year        = {1975}
}

@article{coleman1975,
  author  = {Coleman, Meri and Liau, Ta Lin},
  title   = {A Computer Readability Formula Designed for Machine Scoring},
  journal = {Journal of Applied Psychology},
  volume  = {60},
  number  = {2},
  pages   = {283--284},
  year    = {1975}
}

@book{gunning1952,
  author    = {Gunning, Robert},
  title     = {The Technique of Clear Writing},
  publisher = {McGraw-Hill},
  year      = {1952}
}

@techreport{smith1967,
  author      = {Smith, Edgar A. and Senter, R. J.},
  title       = {Automated Readability Index},
  institution = {Aerospace Medical Research Laboratories},
  number      = {AMRL-TR-66-220},
  year        = {1967}
}

\end{document}